\documentclass[aps,prb,twocolumn,superscriptaddress,showpacs,floatfix,10pt]{revtex4-1}
\usepackage{amsmath}
\usepackage{amsfonts}
\usepackage{graphicx}
\usepackage{color}
\usepackage{epstopdf}
\usepackage{natbib}
\usepackage{xfrac}
\usepackage{enumitem}
\usepackage[english]{babel}
\usepackage{booktabs}

\usepackage{float} 
\usepackage{subfigure}

\begin{document}
\title{Removal of spectral distortion due to echo for ultrashort pulses propagating through multilayer structures with thick substrate}

\author{Yingshu Yang}
\affiliation{School of Physical and Mathematical Sciences, Nanyang Technological University, Singapore, Singapore}
\author{Stefano {Dal~Forno}}
\affiliation{School of Physical and Mathematical Sciences, Nanyang Technological University, Singapore, Singapore}
\author{Marco Battiato}
\email{marco.battiato@ntu.edu.sg}
\affiliation{School of Physical and Mathematical Sciences, Nanyang Technological University, Singapore, Singapore}

\date{\today}

\begin{abstract}
Given the wide range of applications of time-domain spectroscopy, and particularly THz time-domain spectroscopy, the modelling of a probe pulse propagating through a multilayered structure is often required. Due to the fact that the multilayers are usually grown on a substrate much thicker than the other layers, the transmission of a probe pulse includes a series of echo pulses caused by the multiple reflections at the substrate interfaces. Experiments often measure the time profile and construct the transmitted spectrum only from the first transmitted pulse. Due to the fact that typical substrates lead to times of crossing comparable to the spectral bandwidth, the first transmitted pulse's spectrum and the full transmitted spectrum can be importantly different. It is therefore important to theoretically model the transmission without the echo, to be able to directly compare with experimental results. Here we propose a method to elegantly and easily theoretically remove the echo from the transmission spectrum, without Fourier transforming the signal twice. The spectrum of the transmitted pulse without echo will be produced analytically without additional numerical steps, reducing computational time and program complexity when dealing with larger scale computing.

\end{abstract}

%\keywords{}

\maketitle

In the field of material sciences, researchers are mostly aiming at developing new kinds of materials with specific or customized characteristics using newly designed complex structures. These complex structures are usually multilayered system\citep{geim_van_2013,jariwala_emerging_2014}. These specially tailored multilayers are key components to many different devices that have been widely used in modern technology \citep{diroll_optical_2019}. They are also very important components of photodetectors, transistors, sensors, photo-voltaic cells \citep{vattikuti_chapter_2018} as well as THz emitters \citep{kampfrath_terahertz_2013,wang_ultrafast_2018,cheng_far_2019}.

To study the properties of these specially designed multilayers, an electromagnetic probe in different range of frequencies (for example, the optical range, RF range, THz range and so on) is usually used. Different probes give access to, for instance,  thermal properties of the materials, such as the electron-phonon and phonon-phonon interactions \citep{benisty_intrinsic_1991,bozyigit_soft_2016} as well as the transport \citep{kim_thermal_2006,cahill_nanoscale_2002,vineis_nanostructured_2010,ong_surface_2013}. %These information are significant due to the wide-ranged desire of implementations of the materials in various applications covering different scientific fields from physics to chemistry to material science and even electronic engineering \citep{diroll_optical_2019}.
However, the thickness of the materials used to construct these kinds of multilayers are usually down to nanometer scale. Hence, to hold and stabilize these materials, they are generally grown on another layer, called substrate, with a thickness much larger than the active layers. 

When performing time-domain THz spectroscopy \cite{beard_terahertz_2002,baxter_2011,kampfrath_2013} on a multilayer, a THz probe pulse is sent through it. The transmitted pulse is not measured on a spectrometer, but the time profile of the field is directly probed. When the multilayer is held on a substrate, as the pulse propagates through the whole system, echoes are produced by the air-substrate and substrate-active multilayer interface reflections. Experimentally only the first transmitted pulse is usually relevant, and very often the following echo pulses are not analysed and not even measured. However standard theoretical approaches to the propagation of electromagnetic waves through multilayers do not distinguish those pulses and produce the full spectrum of the transmitted radiation \cite{Yeh_optical_waves_2005,Katsidis_2002,yasuda_measurement_2008}. This is usually not a problem, unless the time it takes the pulse to traverse the substrate is comparable with the pulse timewidth: in this case the full spectrum is not directly comparable with experimental results that measure only the first transmitted pulse. Therefore any theoretical treatment cannot overlook this issue and must provide the spectrum of the first pulse only and not that of the full train of pulses. 

A straightforward way to remove the echoes is by inverse Fourier transforming in time the theoretical transmitted spectrum, then erasing the data of the echoes manually and Fourier transform the pulse in time domain back to a spectrum without echo. However, this process will introduce two additional Fourier transform processes and a data-erasing process. These additional processes will be a waste of time and introduce further complexity to any numerical solver. Another approach is to simply assume the substrate as semi-infinite, yet this does not represents the real geometry.\cite{Yeh_optical_waves_2005} To our knowledge, there have been no attempt at addressing the problem of the removal of the echo analytically in real geometries. Notice that for continuous wave (CW) excitations, echos are part of the signal and should not be removed. As we will show later, removing the echo leads to the disappearance of periodic peaks and valleys in the spectrum of the transmission. These features are often absent even in CW experiments, yet for very different reasons. The propagation through the substrate might be incoherent, and accounting for this effect allows for a spectrum that more closely reproduces CW experiments \cite{Katsidis_2002,Lubberts_1981,Centurioni_2005,Santbergen_2013}.  Different authors attempt some suitable averaging procedure to obtain CW spectra including incoherence.\cite{Centurioni_2005,Troparevsky_2010} We however warn that, even if removing the echo and  describing the incoherence of transmission through the substrate modify the spectrum in ways that might look qualitatively similar at a first glance, the two physical effects are different and, therefore, the resulting spectra are not the same.

\begin{figure}[tb]
    \centering
    \includegraphics[width = 0.5\textwidth]{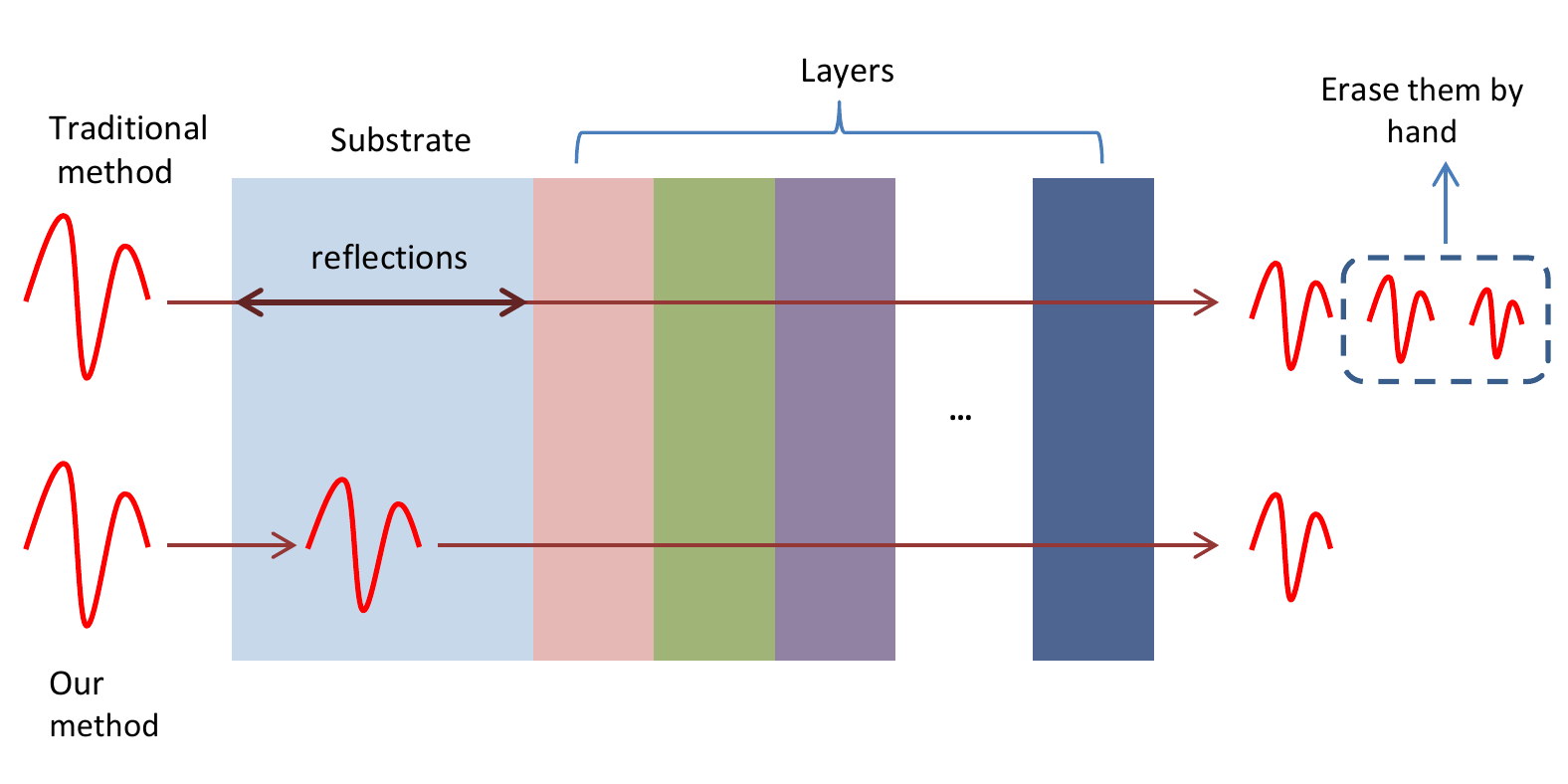}
    \caption[The comparison of the traditional removal of the echo method and our method.]{The comparison of the traditional removal of the echo method and our method.}
    \label{fig:method}
\end{figure}

We propose here an analytical way of removing the echo without the additional steps mentioned above, as shown in Fig.~\ref{fig:method}. Rather than solving the transmission through the whole multilayer, we first calculate the transmission through the substrate and analytically extract the first pulse that  crosses the substrate/active layers interface, while discarding the further propagation that will give rise to echo pulses. We then propagate the mentioned pulse through the active multilayer (now including all multiple reflections) to obtain the transmitted wave. Our method provides an elegant analytical expression of the final transmitted pulse, without the need of avoidable numerical steps, as well as being numerically stable over a huge range of layers thicknesses.

\section{Single layer treatment}
Before considering the echo removal for the whole multilayer system, let us recapitulate the standard transfer matrix approach. We begin with a single layer treatment with orthogonal incidence, where the Maxwell's equations (for one of the two modes) reduce to 
\begin{equation}
  	\partial_z E\left[z,t\right]=  \mu\, \partial_t H\left[z,t\right] 
\end{equation}
\begin{equation}
  	\partial_z H\left[z,t\right]= \epsilon\, \partial_t E\left[z,t\right]
\end{equation}
where $E$ is the electric field, $H$ the magnetising field, $z$ the coordinate running orthogonal (from left to right) to the multi-layer surfaces,  $t$ the time, $\epsilon$ and $\mu$ are the permittivity and the permeability, respectively, of the isotropic material.
As the field within one layer can be represented as the superposition of a left-propagating and right-propagating wave \citep{born_principles_2013} with wave number $k$, and following from the Maxwell equations, we can write
\begin{align}
    E[z,t] &= f^{>}[t]e^{ikz}+f^{<}[t]e^{-ikz},\\
    H[z,t] &= \sqrt{\frac{\epsilon}{\mu}}f^{>}[t] e^{ikz} - \sqrt{\frac{\epsilon}{\mu}}f^{<}[t] e^{-ikz}.
\end{align}
where $k=\omega \sqrt{\mu \epsilon} $. So that if we consider the Fourier components of the time Fourier transform (FT), then the general solution is given by
\begin{equation}
	\mathcal{FT}\left(\begin{bmatrix} E\left[z,t\right]\\ H\left[z,t\right]\end{bmatrix},t\right) =  \begin{bmatrix} E\left[\omega,z\right]\\ H\left[\omega,z\right]\end{bmatrix} =  a\left[ \omega,z \right] \begin{bmatrix} f^{>}\left[ \omega\right] \\ f^<\left[ \omega\right] \end{bmatrix}
	\label{FToffield}
\end{equation}
where 
\begin{equation}
	a[\omega,z]= \begin{bmatrix} e^{i \omega \sqrt{\epsilon \mu} z } &e^{-i \omega \sqrt{\epsilon \mu} z} \\ \sqrt{\frac{\epsilon}{\mu}}e^{i \omega \sqrt{\epsilon \mu} z }&{-\sqrt{\frac{\epsilon}{\mu}} e^{-i \omega \sqrt{\epsilon \mu} z}}\end{bmatrix},
\end{equation}
and $f^{>}$ and $f^{<}$ represent the amplitude of the right and left moving waves respectively.
For brevity we will write the Fourier transformed equation (\ref{FToffield}) as,
\begin{equation}
	\bar{F}\left[\omega,z\right]=a[\omega,z]\bar{f}\left[ \omega\right],
\end{equation}
where,
\begin{align}
	\bar{F}\left[\omega,z\right]&=\begin{bmatrix} E\left[\omega,z\right]\\ H\left[\omega,z\right]\end{bmatrix} \\
	\bar{f}\left[ \omega\right] &=\begin{bmatrix} f^{>}\left[ \omega\right] \\ f^<\left[ \omega\right] \end{bmatrix}. 
\end{align}
When obvious from the context the frequency dependence will be omitted.
Therefore, the values of the fields at the two surfaces of a layer of thickness $d$ can be written as
\begin{align}
	\bar{F}\left[0\right]&=a[0]\bar{f} \\
	\bar{F}\left[d\right]&=a[d]\bar{f},
\end{align}
so that,
\begin{equation}
	\bar{F}\left[d\right]=M[d]\bar{F}\left[0\right]
\end{equation}
with 
\begin{equation}
    \begin{aligned}
	M[d]&=a[d](a[0])^{-1}\\
	&= \begin{bmatrix} \cos(\omega\sqrt{\epsilon \mu}d) & i\sqrt{\frac{\mu}{\epsilon}} \sin(\omega\sqrt{\epsilon \mu}d)\\
	i\sqrt{\frac{\epsilon}{\mu}} \sin(\omega\sqrt{\epsilon \mu}d)&\cos(\omega\sqrt{\epsilon \mu}d)  \end{bmatrix} .  
    \end{aligned}  
\end{equation}

\section{Multilayer treatment}
\begin{figure}[!h]
\centering
\includegraphics[scale=0.7]{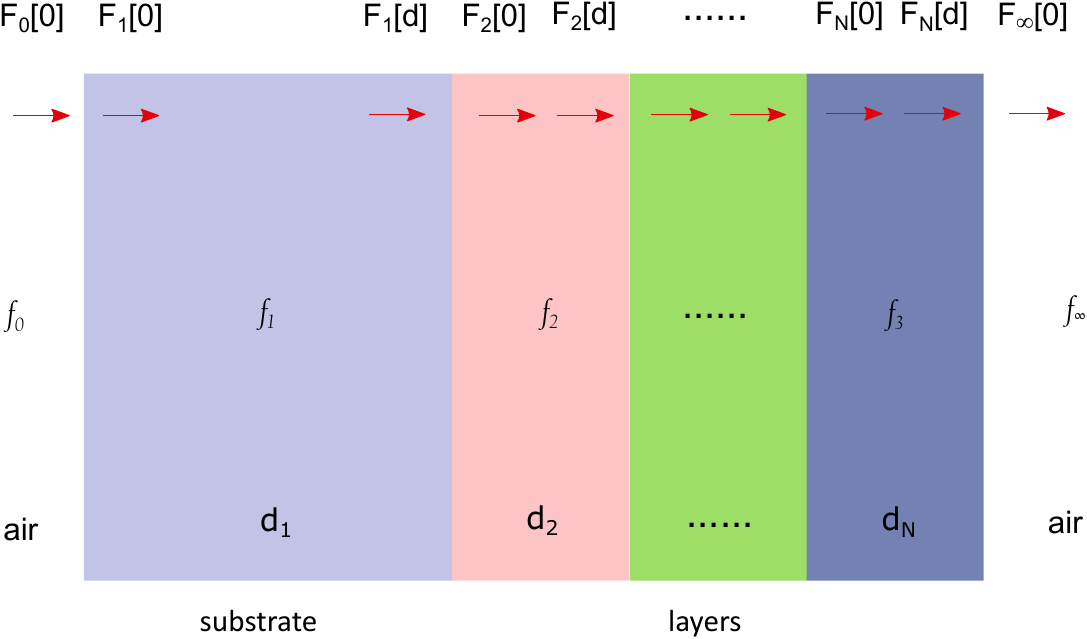}
\caption{The general description of the fields in the multi-layer system. F is the field vector, f is the right and left propagating field amplitude vector and d is the thickness of each layer.}
\label{fig.multilayersystem}
\end{figure}

The transmission across a multilayer is computed requiring that the fields at the interfaces have to be continuous. We assume the cases when the multilayer is sandwiched by air or vacuum, as shown in Fig.~\ref{fig.multilayersystem}, since it is the most common case (the formulas are easily generalised). We will use subscripts to denote matrix and vector properties belonging to a given layer. In particular, the semi-infinite left air layer will be denoted with $0$, the layers of the multilayer will be denoted with increasing numbers from left to right and the semi-infinite air layer on the right will be denoted with $\infty$. A local axis for the $z$ coordinate in each layer, with origin on the left surface (implying that the coordinate of the right surface will be the thickness of the layer) will be used. The only exception is the semi-infinite left air layer where the origin is set at the right surface (since the other surface is at $-\infty$). 
Fig.~\ref{fig.multilayersystem} shows a general case of fields within a multilayer system.
We can then write the following equations enforcing field continuity at the interfaces
\begin{equation}
\begin{split}
	\bar{F}_0\left[0\right]&=a_{0}[0]\bar{f}_0 \\
	\bar{F}_1\left[0\right]&=a_{1}[0]\bar{f}_1  \\
	\bar{F}_1\left[d_1\right]&=a_{1}[d_1]\bar{f}_1 \\
	...\\
	\bar{F}_N\left[0\right]&=a_{N}[0]\bar{f}_N  \\
	\bar{F}_N\left[d_N\right]&=a_{N}[d_N]\bar{f}_N \\
	\bar{F}_{\infty}\left[0\right]&=a_{\infty}[0]\bar{f}_{\infty}.
\end{split}
\end{equation}
By imposing that (with the only exception at the  interface between air and the first layer)
\begin{equation}
	\bar{F}_j\left[d_j\right]=\bar{F}_{j+1}\left[0\right]
\end{equation}
we can obtain all the fields and the amplitudes of the propagating waves. In particular we obtain 
\begin{equation}
	\bar{f}_{\infty} =\left(a_{\infty}[0]\right)^{-1} \left(\prod_{j=N}^{1} M_j[d_j] \right) a_{0}[0]\bar{f}_{0} = T_{[0,\infty]}\bar{f}_{0}
\end{equation}
where we draw the attention to the inverse order in the multiplication. And the $T_{[0,\infty]}$ matrix here is the transfer matrix of the whole multilayer system.

\section{Transmission and reflection}
A transmission experiment is easily modelled by requiring that there is no left propagating wave in the rightmost air layer. This is achieved by solving the system
\begin{equation}
	\begin{bmatrix} f_{\infty}^>\\ 0\end{bmatrix} = T_{[0,\infty]} \begin{bmatrix} f_{0}^>\\  f_{0}^<\end{bmatrix}.
\end{equation}
where $f_{0}^>$ represents the incoming, $f_{0}^<$ the reflected, and $f_{\infty}^>$ the transmitted wave respectively.
The transmission coefficient $t$ is defined as the ratio between the amplitude of the electric field due to the right going wave in the right air layer and the amplitude of the electric field due to the right going wave in the left air layer. The reflection $r$ coefficient is analogously defined as the ratio between the amplitude of the electric field due to the left going wave in the left air layer and the amplitude of the electric field due to the right going wave in the left air layer\citep{born_principles_2013}. In the considered case when the multilayer is surrounded by air on both sides the two coefficients simplify to the ratio of the corresponding amplitudes of the waves as
\begin{equation} \label{eq:transm_coeff}
 	t_{[0,\infty]} =\frac{f_{\infty}^>}{f_{0}^>}= \frac{T_{[0,\infty],11}  T_{[0,\infty],22}-T_{[0,\infty],12}\,T_{[0,\infty],21}}{T_{[0,\infty],22}}  
\end{equation}
\begin{equation}
    r_{[0,\infty]}=\frac{f_{0}^<}{f_{0}^>}= - \frac{T_{[0,\infty],21}}{T_{[0,\infty],22}}    
\end{equation}
where the further subscripts represent matrix elements. Thus, we are able to write out the transmitted and reflected wave as,
\begin{equation}
 	f_{\infty}^>=t_{[0,\infty]} \; f_{0}^> ,
\end{equation}
and
\begin{equation}
    f_{0}^<= r_{[0,\infty]} \; f_{0}^> .
\end{equation}

\section{Removal of the echo within the substrate in the transmission}
For the removal of the echo, we consider a sample grown on a layer, the substrate, much thicker than the other ones. We will use the subscript $S$ for the substrate layer, and  let the layer indices start at 1 for the first layer after the substrate.

To remove the echo, we first need to identify the first pulse travelling through the substrate, triggered by the incoming pulse on the left-hand air side and exclude the wave reflected at the substrate/layer 1 and all its multiple reflections. This can be simply constructed by imagining a sample formed only by the air/substrate interface. In that configuration the right propagating wave in the substrate can be written as,
\begin{equation}
 f_{S}^{>*} = t_{\left[0,S\right]} f_{0}^>.
\end{equation}
where $t_{\left[0,S\right]} $ is calculated from Eq.~\ref{eq:transm_coeff} with the transfer matrix $T_{[0,S]}$ constructed as
\begin{equation}
    T_{[0,S]} = (a_{S}[0])^{-1} a_{0}[0].
\end{equation}
giving 
\begin{equation}
 	t_{[0,S]} = \frac{T_{[0,S],11}  T_{[0,S],22}-T_{[0,S],12}\,T_{[0,S],21}}{T_{[0,S],22}}  
\end{equation}
Notice that $f_{S}^{>*} $ indeed is the first propagating pulse through the substrate (excluding the reflection pulse and all the subsequent ones due to multiple reflections within the substrate) even when the full sample is considered.

Such pulse will travel through the substrate and reach the interface with the rest of the sample. At that interface it will then undergo reflection, as well as propagation through the active part of the sample (inside which we need to consider multiple reflections). We can treat $f_{S}^{>*} $ as an incoming wave for the rest of the multilayer and solve
\begin{equation}
	\begin{bmatrix} f_{\infty}^{>\scriptstyle{\text{no echo}}}\\ 0\end{bmatrix} =T_{[S,\infty]} \begin{bmatrix}  f_{S}^{>*}\\  f_{S}^{<*}\end{bmatrix} = T_{[S,\infty]} \begin{bmatrix} t_{\left[0,S\right]} f_{0}^>\\  f_{S}^{<*}\end{bmatrix}
\end{equation}
where the transfer matrix $T_{[S,\infty]}$ is
\begin{equation}
    T_{[S,\infty]} =  (a_{\infty}[0])^{-1} \left(\prod_{j=N}^{1} M_j[d_j] \right) a_{S}[d_S],
\end{equation}
where we remind that $j=1$ now refers to the first layer after the substrate.
As we are interested only in the transmitted pulse, we obtain the overall transmission coefficient for the no-echo case as
\begin{equation} \label{eq:noecho}
    t_{[0,\infty]}^{\scriptstyle{\text{no echo}}} = t_{[0,S]}t_{[S,\infty]},
\end{equation}
where
\begin{equation}
 	t_{[S,\infty]} = \frac{T_{[S,\infty],11}  T_{[S,\infty],22}-T_{[S,\infty],12}\,T_{[S,\infty],21}}{T_{[S,\infty],22}}. 
\end{equation}

\section{Simulation results}

We here show numerical results obtained using the technique introduced above for the propagation of a pulse. We show results for pulses in the THz frequency domain, but the approach above is general. We choose as an example a multilayer structure composed by a sapphire ($\rm Al_2O_3$) substrate and an aluminum ($\rm Al$) layer covered by a silica ($\rm SiO_2$) layer with 0.5mm, 0.5nm and 4nm thicknesses respectively (notice the difference in the units). Sapphire is a commonly used substrate in the THz range due to its low absorption. The relative permittivities of the sapphire and silica are nearly frequency independent on the THz range and are calculated using\cite{wooten_optical_2013} 
\begin{equation}
    \epsilon_r = n^2-\kappa^2 + i 2n\kappa,
\end{equation}
where the refractive index $n$ ( 3.31 for sapphire and 1.98 for silica \cite{sanjuan_optical_2012}). The $\kappa$ is the extinction coefficient taken as 0.002 for sapphire\cite{palik_handbook_1998} and 0.4 for silica \cite{kitamura_optical_2007}.  The optical properties of aluminum  are instead  frequency-dependent and the data on its dependence is taken from Ref.~\onlinecite{hagemann_optical_1975}.\par

\begin{figure}[tb]
    \centering
    \includegraphics[width = 0.48\textwidth]{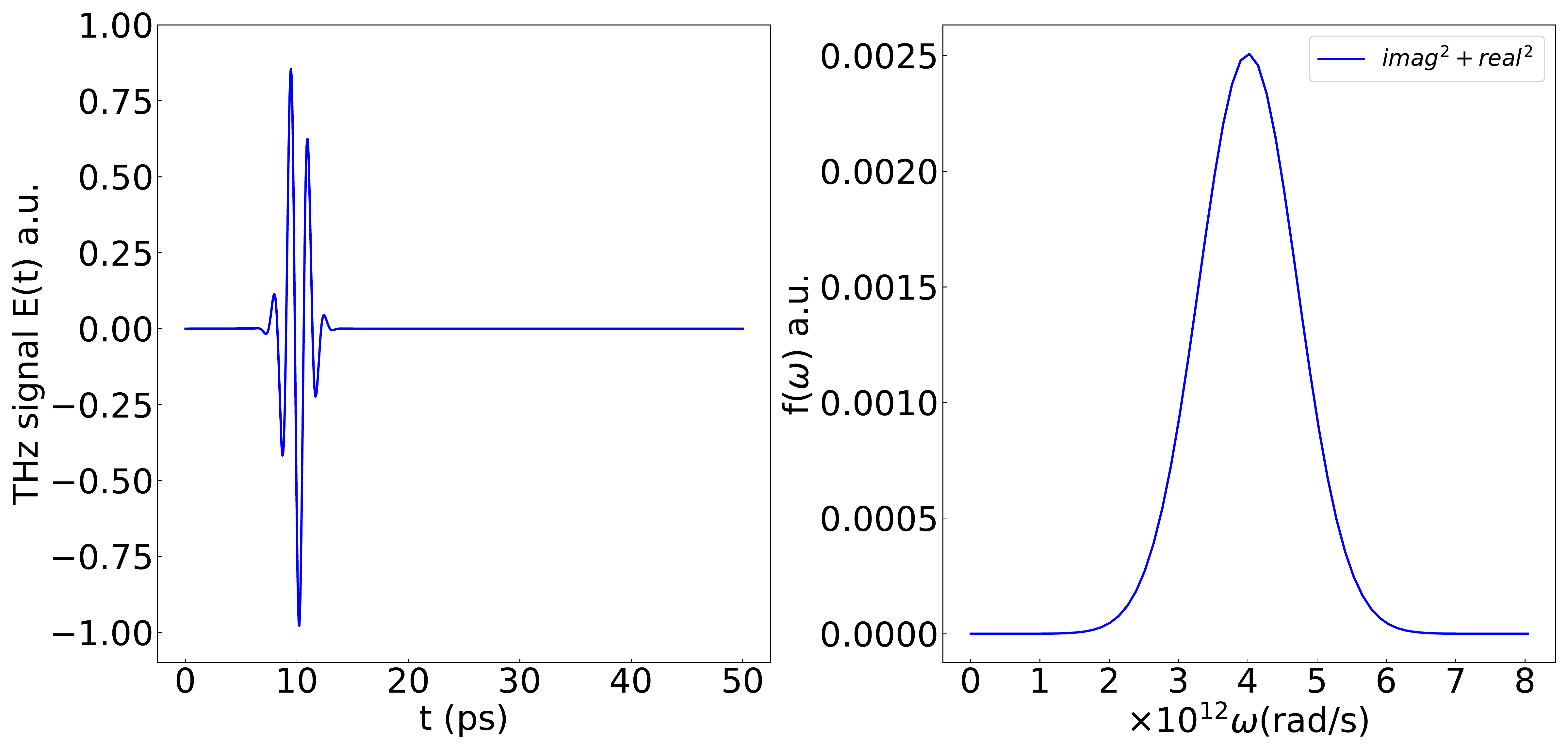}
    \caption[The incident THz pulse in time domain (left) and in frequency domain (right).]{The THz pulse in time domain (left) and in frequency domain (right).}
    \label{fig:inputandspectrum}
\end{figure}

We suppose the multilayer to be probed by a THz pulse coming from the substrate side, for which, as an example, we use a simple Gaussian temporal profile with a specific central frequency (however the developed code can handle any temporal profile)
\begin{equation}
    F(t) = e^{\frac{-0.5(t-\mu)^2}{\sigma^2}}  \cos{(\omega t)},
    \label{fun:pulsefunction}
\end{equation}
where $\mu$, the starting time position of the pulse (taken as $ 1 \times 10^{-11}$ s), and $\sigma$, the pulse time length (taken as $ 1\times 10^{-12}$ s). The time window for $t$ is taken from 0 to $5 \times 10^{-11}$ s with 1000 steps. The central frequency $\omega$ is $ 4\times 10^{12}$ rad/s. The input THz pulse and its Fourier Transform (FT) are shown in Fig.~\ref{fig:inputandspectrum}.

\begin{figure}[tb]
    \centering
    \includegraphics[width = 0.48\textwidth]{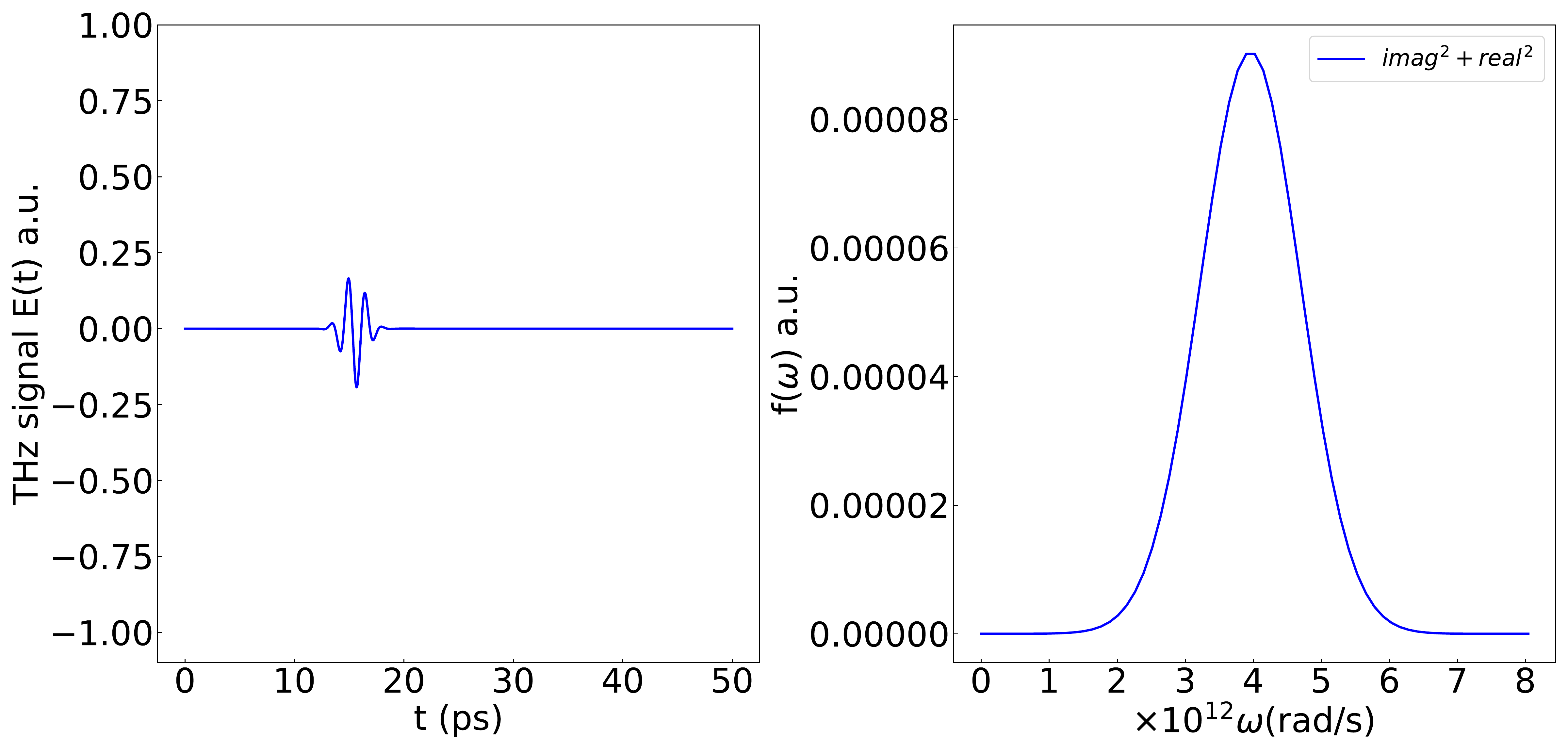}
    \caption[The transmitted THz pulse without echo in time domain (left) and in frequency domain (right).]{The transmitted THz pulse without echo in time domain (left) and in frequency domain (right).}
    \label{fig:transnoecho}
\end{figure}

The full transmitted THz wave through the system is shown in Fig.~\ref{fig:transwaveandspectrum} below. 
 \begin{figure}[tb]
    \centering
    \includegraphics[width = 0.48\textwidth]{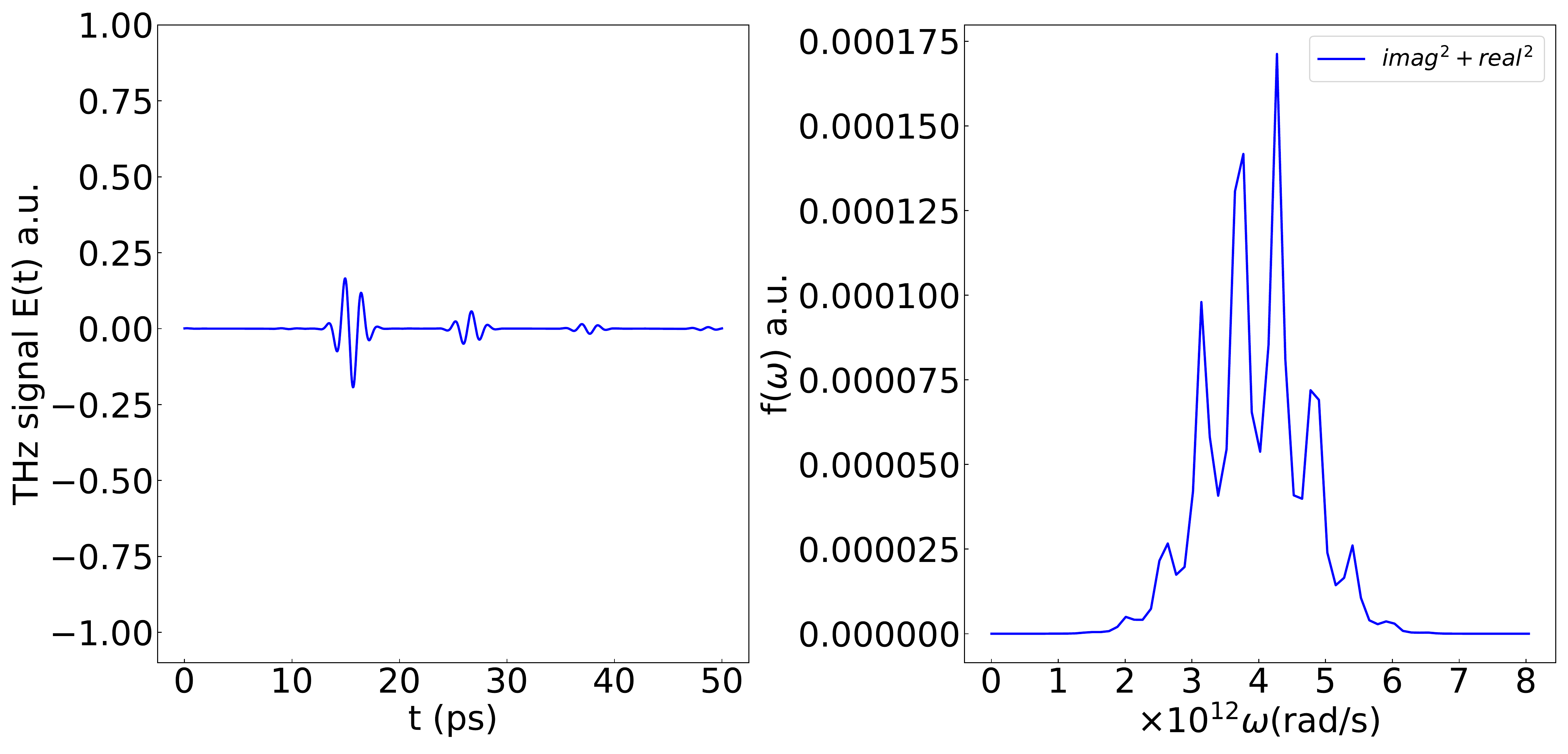}
    \caption[The transmitted THz pulse in time domain (left) and in frequency domain (right).]{The transmitted THz pulse in time domain (left) and in frequency domain (right).}
    \label{fig:transwaveandspectrum}
\end{figure}
As we can see, the transmitted wave in time domain is a main transmitted pulse with a series of echoes. The spectrum of this series of transmitted pulses contains strong oscillations with frequency as shown in the right hand side of Fig.~\ref{fig:transwaveandspectrum}. %These wiggles will hide the information we need when dealing with comparisons with the experimental results. Thus, it became significant to remove the effect of the echoes to the spectrum.
When instead we remove the echo applying the presented method, the final transmitted wave of the THz pulse and its spectrum will be as shown in Fig.~\ref{fig:transnoecho}.
This second spectrum is the one that is usually measured in experiments. One can notice the stark difference between the spectrum including all the echo pulses and the one obtained by instead removing them. As they are hardly comparable, it is clear that any theoretical method aiming at describing THz experiments in the presence of substrates has to remove the echo pulses to be able to make meaningful comparisons with experiments, otherwise the strong oscillations of the full spectrum will not allow for a comparison with experimental spectra.\par

\begin{figure}[tb]
\centering  
\subfigure[]{
\label{Fig.sub.1}
\includegraphics[width=0.48\textwidth]{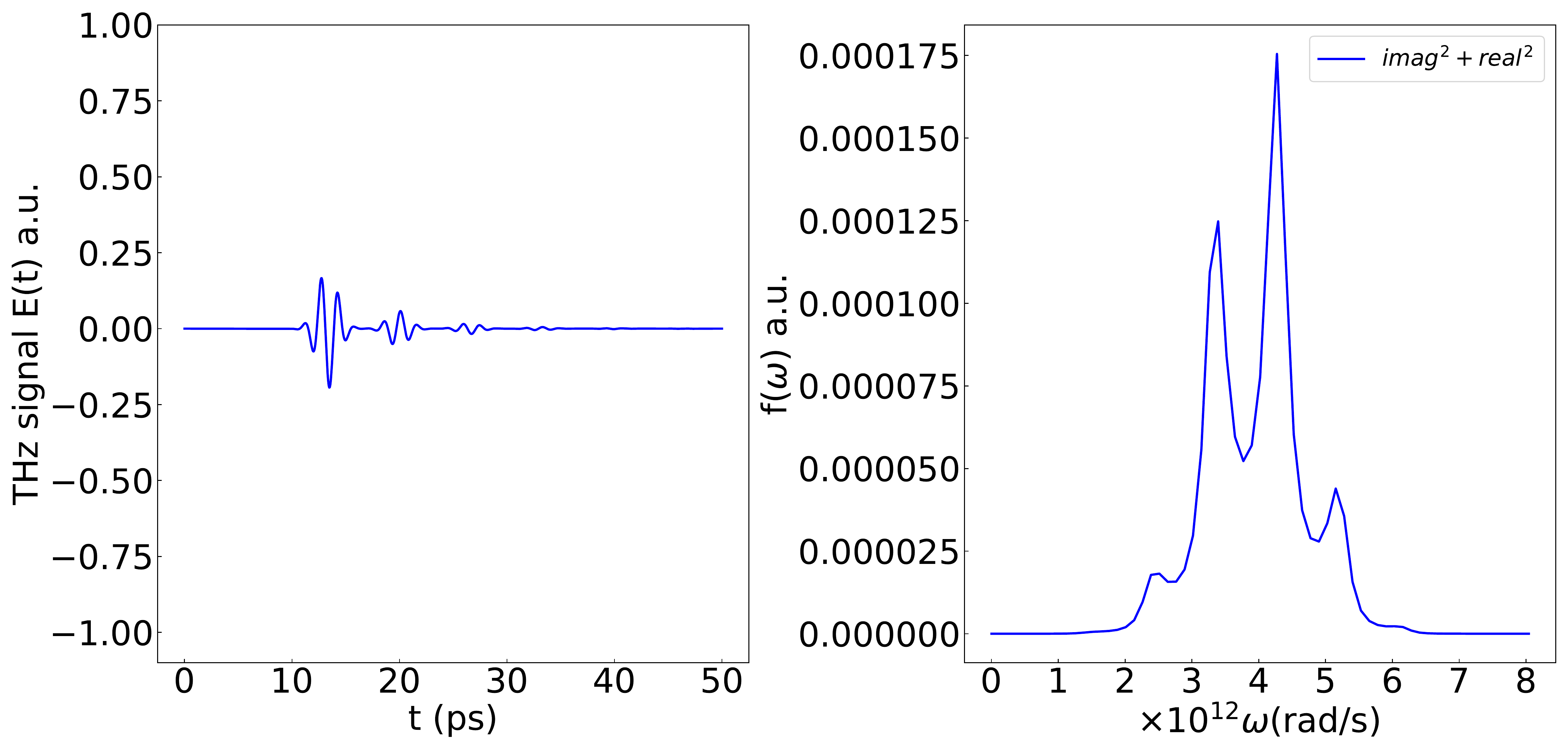}}
\subfigure[]{
\label{Fig.sub.2}
\includegraphics[width=0.48\textwidth]{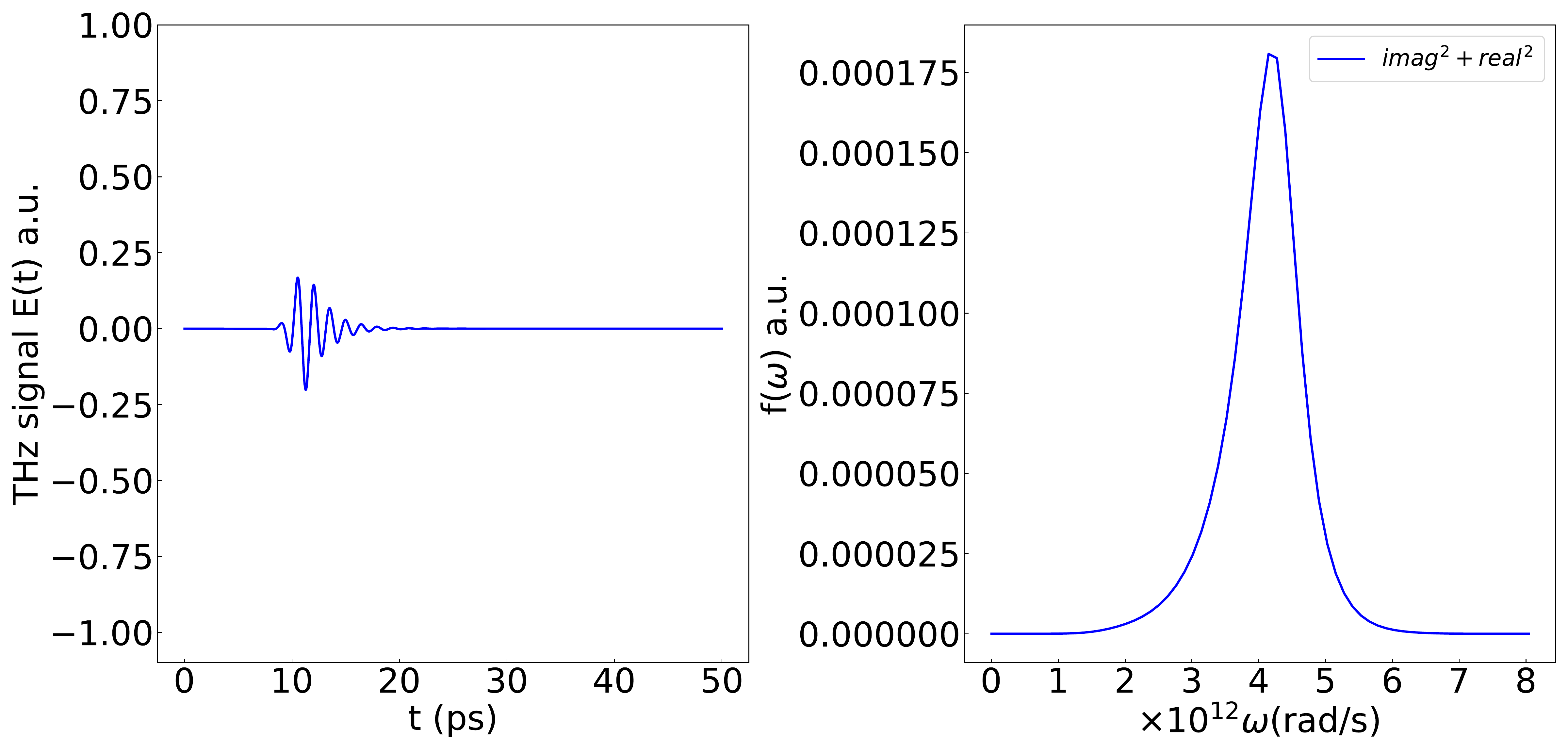}}
\caption{The transmitted external THz pulse through a multilayer system with decreasing thicknesses of substrate without removing the echo. (a) Transmitted wave with 0.0003m thick substrate. (b)Transmitted wave with 0.0001m thick substrate. }
\label{diffsub}
\end{figure}

\begin{figure}[tb]
\centering  
\subfigure[]{
\label{sub1}
\includegraphics[width=0.45\textwidth]{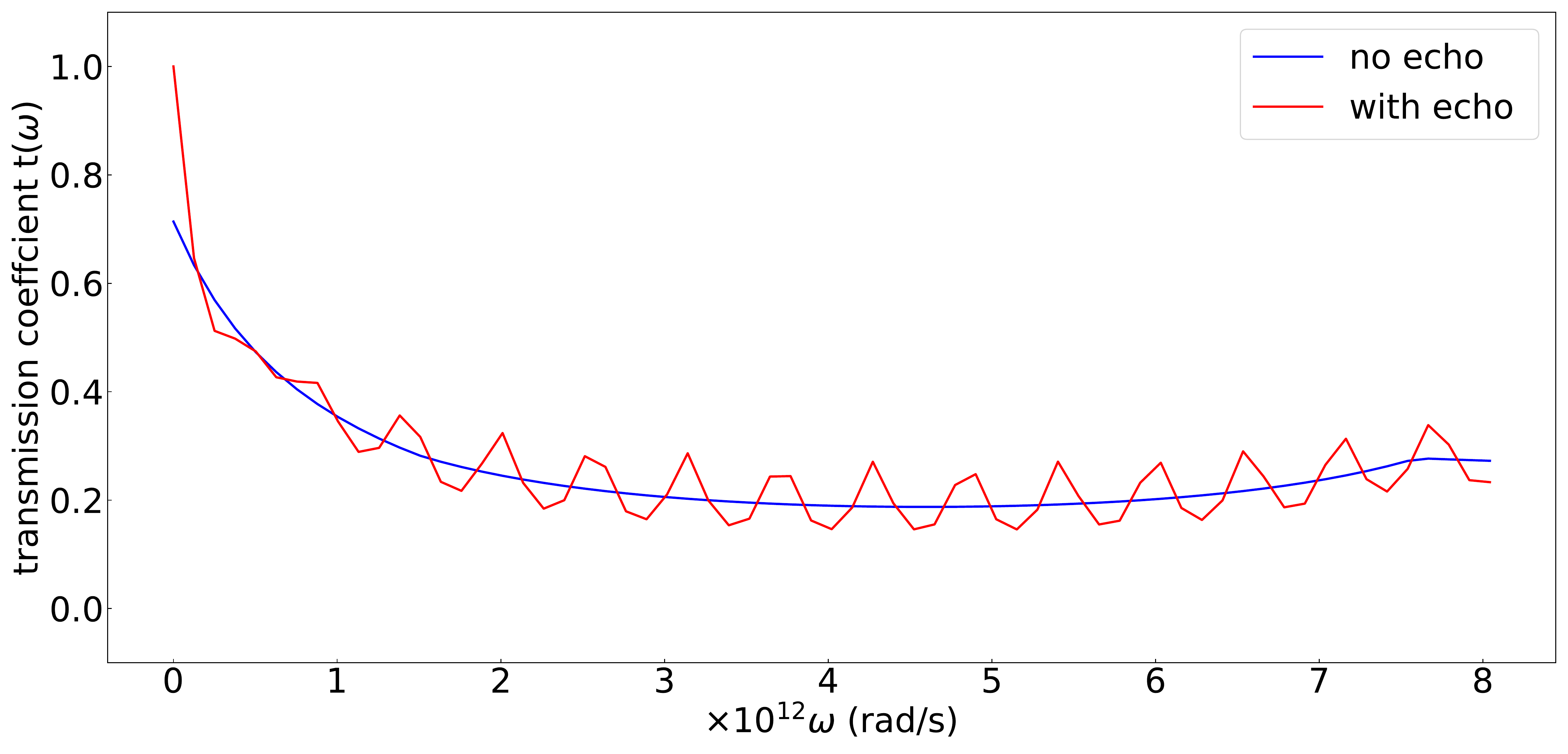}}
\subfigure[]{
\label{sub2}
\includegraphics[width=0.45\textwidth]{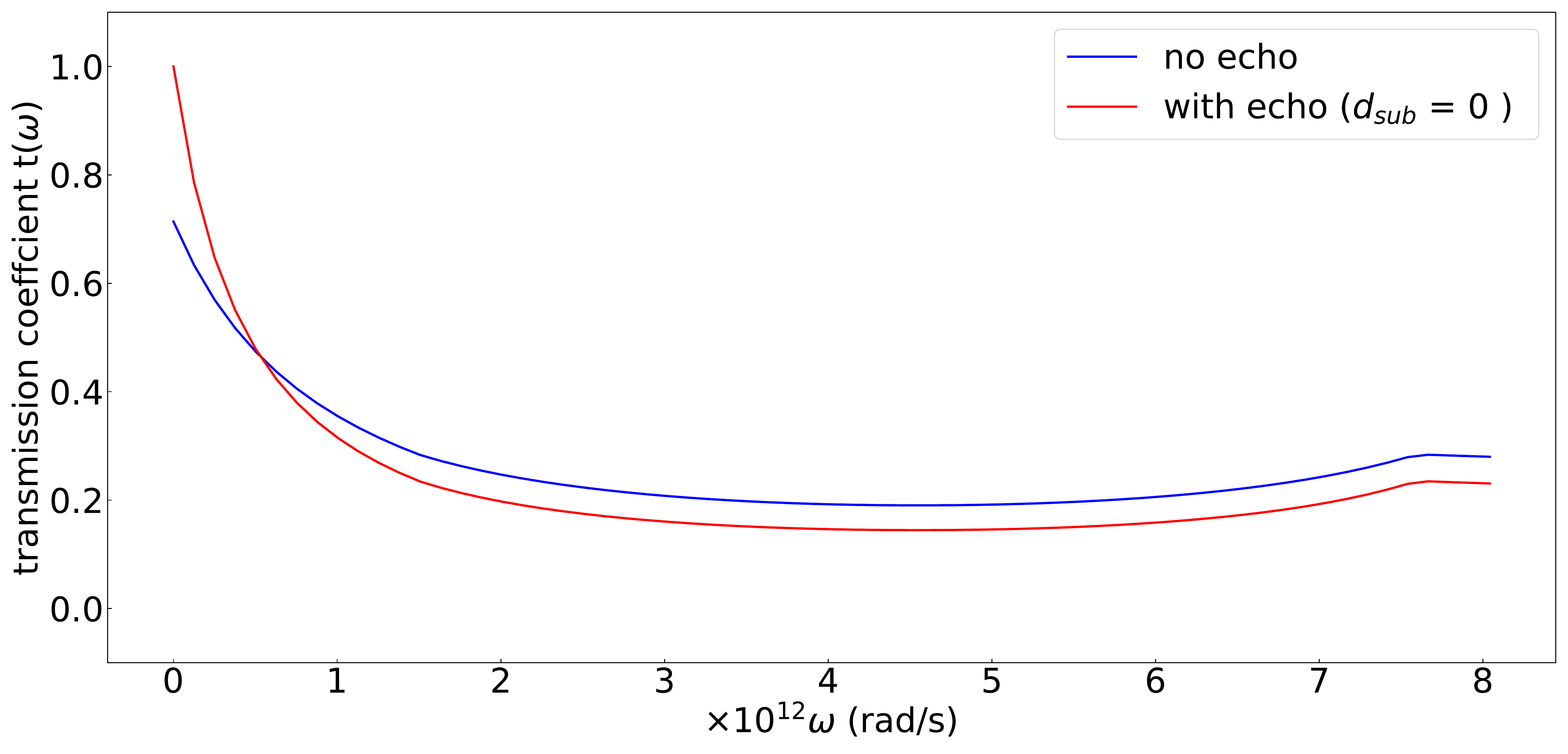}}
\caption{The transmission coefficient for corresponding frequencies before and after removing the echo.(a)Substrates thickness d = 0.5mm.  (c) Substrate thickness d = 0mm.}
\label{fig:tran_coeff}
\end{figure}

It is however important to remind that removing the echo does not always make sense. When the substrate is thin enough, the train of echo pulses will be overlapping with the main transmitted pulse. It is therefore experimentally impossible to avoid measuring the echo signal, and, similarly, it is theoretically meaningless to do so. As we can see from the Fig.~\ref{diffsub}, if we reduce the thickness of the substrate from 0.5mm used in the beginning to 0.3mm, and then to 0.1mm, the echoes coming after the first main wave will be closer to each other and eventually interfere with each other when the substrate is thin enough (bottom of Fig.~\ref{diffsub}). Even though equation \ref{eq:noecho} provides an expression for the transmitted signal without echo pulses generated within the substrate, when the substrate is too thin to form a clear time delay between the transmitted pulses, it is meaningless to remove the echo. 
This can also be seen from the transmission coefficient for corresponding frequencies before and after removing the echo, as shown in Fig.~\ref{fig:tran_coeff}. We can see from Fig.~\ref{sub1} that as the echo is removed, the transmission coefficient spectrum is importantly altered. As experiments will most probably measure all the pulses and extract their spectrum, removing the echo in that case will lead to strong disagreements. 

A further special interesting case of when using Eq.~\ref{eq:noecho} is wrong is when the substrate thickness is set to 0. Eq.~\ref{eq:noecho} and Eq.~\ref{eq:transm_coeff} do not have the same limit for vanishing substrate thicknesses (again because Eq.~\ref{eq:noecho} is not valid for small substrate thicknesses). From Fig.~\ref{sub2} we can see that, the spectrum before and after removing the echo do not overlap. This is a particularly deceiving case, as the two spectra are qualitatively similar.
 
Before proceeding, we also highlight that the numerical approach based on the calculation of the analytic expression has a high sensitivity to the thickness of the layers as it works equally well for a large range of thickness and for samples with layers with thicknesses that vary by several orders of magnitude without giving numerical truncation errors. 

\begin{table}[tb]
\caption{The sequence of the layers and their corresponding thickness.} \label{tab:multi}
\begin{tabular}{c|c|c p{6cm}}
\hline  
                & Before swapping   & After swapping    \\
\hline 
Substrate       & $Al_2O_3 (0.5mm)$ &  $Al_2O_3 (0.5mm)$\\
Layer1          & $Al (0.5nm)$      &  $SiO_2 (4nm)$    \\
Layer2          & $SiO_2 (4nm)$     &  $Al (0.5nm)$     \\
\hline
\end{tabular}
\end{table}

\begin{figure}[tb]
    \centering
    \includegraphics[width = 0.49\textwidth]{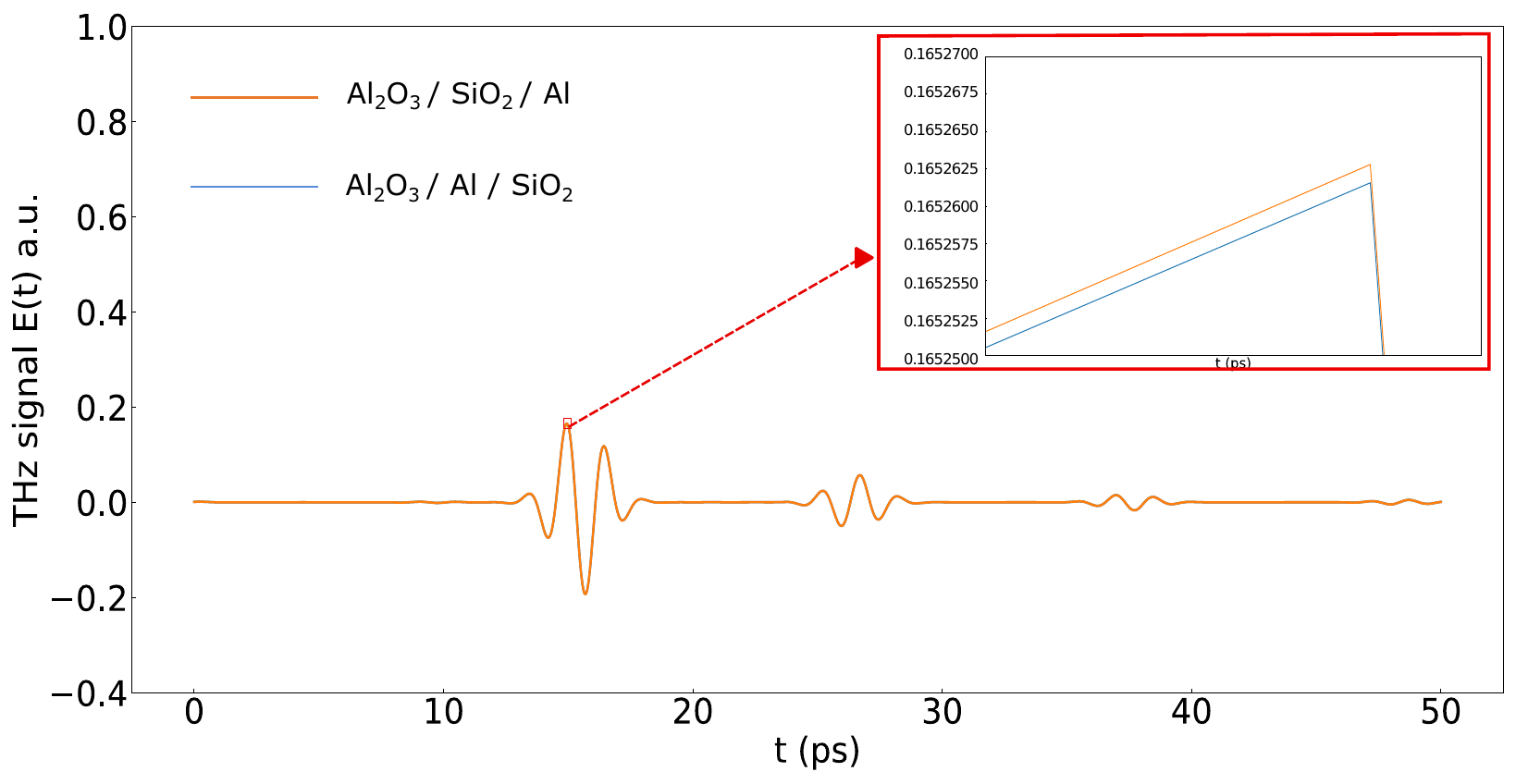}
    \caption[The difference of the transmitted pulse after swapping the layers. Inset: Enlarged peak of the transmitted pulse.]{The difference of the transmitted pulse after swapping the layers. Inset: Enlarged peak of the transmitted pulse.}
    \label{fig:swap}
\end{figure}

As a final example we show the calculated transmitted wave through two multilayers grown on the same substrate, but where the two thin layers have been swapped (see table \ref{tab:multi}). Given that the wavelength of the THz radiation is much larger than the two layers thickness, the transmission is expected to be practically the same, otherwise that would violate the diffraction limit (as a wave cannot resolve the position of an object with a precision higher than its wavelength). In fact, when we swap the layers, we find a very small difference between the transmitted wave before and after swapping the layers.  As we can see from  Fig.~\ref{fig:swap}, the difference between transmitted wave through the $\rm Al_2O_3/ Al/ SiO_2$ set layers and $\rm Al_2O_3/ SiO_2/ Al$ set layers is small compared to the amplitude of the transmitted wave (yet, as expected, it is non zero) of the order of $d/\lambda \approx 10^{-6}$ . 

\section{Conclusion}

In conclusion, we have demonstrated a novel and more effective way of removing the echo caused by the substrate on pulsed radiation propagating through a multilayer system grown on a thick substrate. This allows for a direct comparison with the experimental transmission spectrum of a time dependent probe, where usually the echo pulses are not measured. This method allows for the direct use of an explicit analytic formula and therefore makes it unnecessary to use perform two Fourier transforms and a time filter to obtain the same effect, or to alter the geometry of the experiment.

\bibliography{removeecho}

\end{document}